\documentclass{aa}
\usepackage{psfig}

\begin{document}

\def\pdot {\dot P}
\def\Omdot {\dot \Omega}
\def\ltsima{$\; \buildrel < \over \sim \;$}
\def\lsim{\lower.5ex\hbox{\ltsima}}
\def\gtsima{$\; \buildrel > \over \sim \;$}
\def\gsim{\lower.5ex\hbox{\gtsima}}
\def\msole{~M_{\odot}}
\def\mdot {\dot M}
\def\gr {GRB 030329~}
\def\xte {\textit{Rossi--XTE~}}
\def\xmm  {\textit{XMM--Newton~}}



\title{Late evolution of the X--ray afterglow of GRB 030329}

   \author{
   A. Tiengo\inst{1}$^{,}$\inst{2},
   S. Mereghetti\inst{1},
   G. Ghisellini\inst{3},
   F. Tavecchio\inst{3},
   and G. Ghirlanda\inst{3}
   }


   \institute{Istituto di Astrofisica Spaziale e Fisica Cosmica -- CNR,
              Sezione di Milano ``G.Occhialini'',
          Via Bassini 15, I-20133 Milano, Italy
         \and
             Universit\`{a} degli Studi di Milano,
            Dipartimento di Fisica, v. Celoria 16, I-20133 Milano, Italy
         \and
         INAF-Osservatorio Astronomico di Brera, v. Bianchi 46, I-23907 Merate (LC), Italy
}


\abstract{
The X--ray afterglow of 
the Gamma-Ray Burst 
\gr,  associated to SN2003dh at $z$=0.1685,
has been observed with \xmm 258 days after the burst explosion.
A source with flux
of (6.2$\pm$2.3)$\times10^{-16}$ erg cm$^{-2}$ s$^{-1}$ (0.5-2 keV)
has been detected 
at the GRB position.
This measurement, together with a re--analysis of the 
previous X--ray observations, indicates a
flattening of the  X--ray light curve  $\sim$40 days after the burst. 
This  is in remarkable agreement with the scenario
invoking the presence of two jets with different opening angles.
The wider jet should be responsible for the observed 
flattening due to its transition into the non--relativistic Sedov--Taylor phase.
}

\authorrunning{A. Tiengo et al.}

\maketitle

%

\section{Introduction}

Multifrequency studies of  Gamma--Ray Burst (GRB) afterglows at late times
($\gsim$10 days)  can be used to test
some predictions of the standard, as well as alternative, models.
This has been done so far for a handful of bursts
(e.g., Frail, Waxman \& Kulkarni 2000; Frail et al. 2004).
Thanks to its high flux and low redshift
($z$=0.1685, Greiner et al. 2003; Caldwell et al. 2003), GRB 030329 is  an 
ideal target for such kind of investigations. 

In fact \gr 
had a very large fluence of $\sim$10$^{-4}$ erg cm$^{-2}$ (30--400
keV, Ricker et al. 2003), in the top 1\% of all  detected GRBs,
and its optical transient had magnitude 13 one hour after the
explosion (Peterson \& Price 2003; Torii 2003).
\gr is also the first GRB
unambiguously associated with a supernova (SN2003dh, Stanek et al. 2003;
Hjorth et al. 2003). 
Systematic multiwavelength monitoring of its afterglow (e.g., Lipkin et al. 
2004; Matheson et al. 2003; Tiengo et al. 2003; Sheth et al. 2003) is 
yielding important information for the
understanding of the jet structure, GRB energetics and circumburst
environment.

Here we report on a new \xmm observation  performed 258 days after
the burst in order to study the late evolution of the X--ray afterglow.
We also re--analysed the previous observations to derive in a consistent
way all the X--ray fluxes  in the 0.5-2 keV energy range. This is the range
where the effective area of the \xmm EPIC instrument is highest and
the corresponding fluxes are less affected by uncertainties in the
afterglow spectral shape.

\section{Data analysis and results}

\textit{XMM--Newton} observed the position of \gr starting on December
12, 2003 at 7:30 UT, for an observation length of $\sim$24 hours.
After excluding the time intervals affected by high particle
background, the net exposure times were 68 and 73 ks, respectively in
the PN and MOS cameras of the EPIC instrument (Str\"{u}der et al. 2001; 
Turner et al. 2001).
All the cameras operated in Full Frame mode and with the thin optical
blocking filter.  The data were processed using SAS version 5.4.1.

A faint source at the GRB position is visible, albeit barely,  in 
the 0.5--10 keV images.  The previous
X--ray observations (Tiengo et al. 2003) indicated a rather soft
spectrum and a low absorption, which, combined with
the high effective area of the PN camera at very low energies, results
in a significant fraction of the source counts in the 0.2--0.5 keV
band.  To increase the signal to noise ratio we therefore analyzed the
data over the larger 0.2--10 keV range.  This required an additional
cleaning procedure\footnote{ we processed the PN data using an
experimental version of the task ``epchain'' included in a development
version of the SAS software} to reduce the instrumental noise, which
in the PN detector increases very rapidly below $\sim$0.4 keV.  The
resulting image is shown in Fig.1, where the small solid circle
indicates the \gr position.

The estimate of the afterglow flux requires particular care, due to
the proximity ($\sim$30$''$) of an X-ray emitting AGN, which in this
observation was much brighter than the GRB afterglow.  For this reason
we followed different methods to estimate the background and the
fraction of AGN counts falling in the source extraction region.
For the afterglow counts we used a
circular region with radius of 15$''$ centered at the accurately known
GRB position (Taylor et al. 2003).
The total number of counts (i.e. background, AGN contamination, and afterglow)
contained in this region is  148 for the PN and 174 for the sum of the two MOS.

With the first method, we estimated separately the background and the
AGN contamination. The former was derived from a source free region (a
circle of radius 40$''$) in the same CCD chip, while our estimate of
the latter was based on the well known model of the instrumental Point
Spread Function (PSF, Ghizzardi 2002). We measured the net AGN counts
within 15$''$ from the AGN position (572$\pm$26 in the PN and
456$\pm$25 in the two MOS summed).  The PSF model predicts that a
number of counts corresponding to 5\% and 4\% of these values (for the
PN and MOS cameras, respectively) will fall in the 15$''$ radius GRB extraction
region.  Thus we finally obtain an estimate of 50$\pm$19 net afterglow
counts in the 0.2--10 keV energy band (33$\pm$13 in the PN camera
alone). 

An alternative approach consists in directly measuring the background
and AGN contamination based on an appropriate extraction region close
to the afterglow position, as was done in \cite{tiengo}.  In this way 
we obtain 54$\pm$19 net afterglow counts in the 15$''$ GRB extraction 
region (see Fig. 1).

As a further test we developed a maximum likelihood analysis program and  used it 
to model the
region of the 0.2--10 keV PN image containing \gr and the AGN.  We
used the appropriate PSF and let the positions and intensities of the
two sources, as well as the background level, as free parameters.
This yielded an intensity normalization for the afterglow
corresponding to 34 PN counts within a radius of 15$''$. The source
positions, the background level and the AGN intensity derived in this
way were also consistent with the values found with the methods
described above.

In conclusion, since all methods gave consistent results, in
the following we adopt the count estimate obtained with the first method, 
which corresponds to a 2.7$\sigma$ detection.
Since the small statistics does not permit to carry out a detailed spectral
analysis, a spectral model must be assumed to convert
count rates to physical flux units.
The spectra measured in the previous X--ray 
observations  were well fitted by a power law. 
In the first \xte observation  
the  photon index could be tightly 
constrained at $\Gamma$=2.17$_{-0.03}^{+0.04}$. The
following observations, performed by \xmm, gave a strong upper limit on the
neutral hydrogen column density, which cannot differ substantially
from the Galactic  value of N$_{\rm H}$=2$\times10^{20}$ cm$^{-2}$.
We have then verified that an absorbed power law  with $\Gamma$=2.17
and $N_{\rm H}=2\times10^{20}$ cm$^{-2}$ gives an acceptable fit to all the X--ray 
spectra of the \gr afterglow measured up to now. 
We thus assume such a spectrum, which yields  an 
observed flux  (i.e. not corrected for the absorption)
of (6.2$\pm$2.3)$\times10^{-16}$ erg cm$^{-2}$ s$^{-1}$ in the  0.5--2 keV range.

\section{The X--ray afterglow long term evolution}

To derive the long term flux history of the  afterglow in a
consistent way, we have re-analyzed the two \xmm observations of May
2003 using the procedures described above. In particular, we
reprocessed the PN data with the new cleaning algorithm
and verified that the different
ways of estimating the AGN contamination gave consistent results.

As for the December observation, we finally adopt the values based on
the first method: 501$\pm$25 and 400$\pm$25 net afterglow counts for the
first and second observation, respectively. 
The light curve from   the three \xmm observations is well fit by 
a power law decay with  index $\delta$=1.45$\pm$0.2 (Fig. 2).

For a power law spectrum with $\Gamma$=2.17 and $N_{\rm H}=2\times10^{20}$ 
cm$^{-2}$, the above  count rates correspond to 0.5--2
keV observed fluxes of (143$\pm$7)$\times10^{-16}$ and
(79$\pm$5)$\times10^{-16}$ erg cm$^{-2}$ s$^{-1}$, respectively
\footnote{the corresponding values in the  0.2--10 keV range
((351$\pm$17)$\times10^{-16}$ and (194$\pm$12)$\times10^{-16}$ erg
cm$^{-2}$ s$^{-1}$) are slightly different, 
but consistent within the uncertainties, from those
reported in \cite{tiengo}, which were derived from spectral
fitting}.
These flux
estimates do not change significantly if the spectral parameters are
allowed to vary within their 90\% confidence intervals 
(Tiengo et al. 2003).

To draw a complete X--ray light curve containing both the \xmm and
\xte  data, an extrapolation of the X--ray spectrum is
unavoidable. In fact, most of the counts detected by EPIC have
energies below 2 keV, while the PCA instrument of the \xte satellite is
not sensitive to such low energy photons.  We therefore extracted the
0.5--2 keV fluxes of the \xte observations from the best fit
absorbed power law model to each PCA spectrum, obtaining
the values reported in Table 1.
The corresponding
errors take into account the uncertainties on the spectral shape.  The
resulting light curve is shown in Figure 3. 
Fitting it with a single power law yields a time decay with $\delta$=1.78$\pm$0.02
and reduced $\chi^2$ of 2.6.


\begin{table}
\begin{tabular}{crcrcr}
\hline
{\bf Time since burst} & {\bf Satellite} & {\bf 0.5--2 keV flux}\\
{\bf (days)} & & {\bf (10$^{-12}$ erg cm$^{-2}$ s$^{-1}$)}\\
\hline
0.2065 & \xte & 153$\pm$15\\
0.2125 & \xte & 146$\pm$15\\
0.2185 & \xte & 162$\pm$15\\
0.2515 & \xte & 134$\pm$18\\
1.2762 & \xte & 9.3$\pm$3\\
37.3 & \xmm & 0.0143$\pm$0.0007\\
60.9 & \xmm & 0.0079$\pm$0.0005\\
258.3 & \xmm & 0.00062$\pm$0.00023\\
\hline
\end{tabular}
\caption{Fluxes of the afterglow of \gr in the 0.5--2 keV energy range.}
\label{tab:a}
\end{table}

\section{Discussion}

As shown above,  a single power law decay gives a marginally acceptable (1.6\% c.l.)
description of the X--ray afterglow from 5 hours to almost nine months from the burst.

The data presented in \cite{tiengo} were consistent with a  break in the X--ray light curve
at t$\sim$0.45 days. Similarly to the optical
data available at that time, the decay slope changed from $\sim$0.9 to $\sim$1.9.
Note that in \cite{tiengo} we used
fluxes in the 2--10 keV energy band, which
was optimal to represent the observed flux in the PCA range,
but implied a substantial extrapolation of the poorly constrained \xmm spectra.
This resulted in 
large errors on the \xmm 2--10 keV fluxes.
Due to the different energy range used here (0.5--2 keV),
the points at t=37 and 61 days have smaller errors.
They are not consistent with a slope of 1.9 but are
instead fitted by a flatter power law (see Fig. 2).

In the following we give a more complex interpretation of the X--ray data
driven by the two-jet model proposed by \cite{berger}.
Although this is not the unique way to fit the X--ray light curve,
it has the advantage of explaining the complete X--ray data set presented here
as a natural consequence of an already published interpretation of 
the light curves at optical and radio wavelengths.

\begin{figure}
\psfig{figure=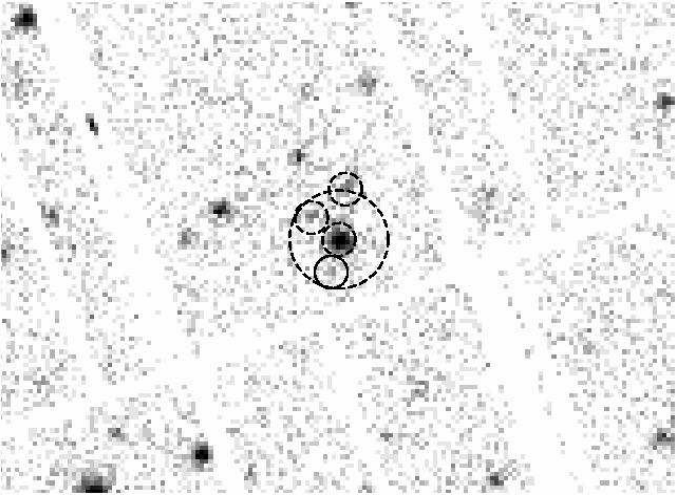,angle=0,width=9cm}
\caption{PN image (0.2--10 keV) of the field of GRB 030329 on December
12--13, 2003. The extraction region for the afterglow (solid circle) and one
of the regions used for the background subtraction are shown.  
The background region is an annulus
centered at the position of the nearby AGN with inner and outer radii of
15$''$ and 45$''$, with the exclusion of the 15$''$ circle around the GRB and
other two weak sources.
}
\label{ima}
\end{figure}

\begin{figure}
\psfig{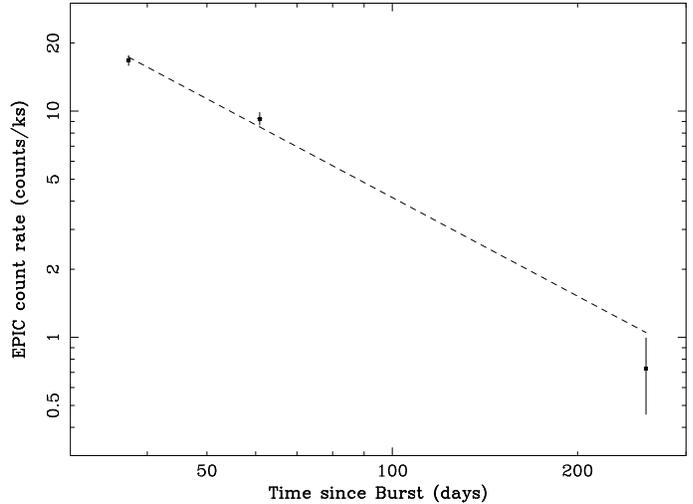}
\caption{
Count rate of the GRB 030329 afterglow in the 0.2--10 keV energy band 
as measured by the EPIC instrument  
(the plotted count rate is from the sum of the PN, MOS1 and MOS2 net counts within 
an extraction region of 15$''$). 
The line is the best-fit power law model ($\delta=1.45\pm$0.2).
}
\label{lcurve}
\end{figure}


On the basis of multifrequency radio data, \cite{berger} 
suggested that GRB 030329 is characterized by a two--component jet,
with different opening angles (5 and 17 degrees).
The wide jet has an intrinsic kinetic energy, corrected for collimation,
of $E=2.5 \times 10^{50}$ erg,
while the narrow jet emits an intrinsic $\gamma$--ray energy
of $E_\gamma=5\times 10^{49}$ erg.
Both jets are assumed to move 
in an interstellar medium with a constant density of $n =1.8$ cm$^{-3}$.  
The initial  part of 
the optical and X--ray afterglow should originate from the
narrow jet, while the wide 
jet should be responsible for the radio flux, the late
X--ray afterglow, and it also explains    
the resurgence of the optical flux after $\sim$1
day (Berger et al. 2003; Sheth et al. 2003; Lipkin et al. 2004).

The isotropic equivalent energies of the two jets are 
$6.6\times 10^{52}$ erg (narrow jet, assuming a kinetic to gamma--rays
conversion factor of $\eta_{\gamma}$=0.2) and
$5.7\times 10^{51}$ erg (wide jet).  
As a consequence, the wide jet should become non--relativistic earlier
than the narrow jet.
The transition to the non--relativistic Sedov--Taylor
phase manifests itself  through a change
in the decaying slope of the flux, which becomes 
$F(\nu, t)\propto t^{-\delta_{NR}}$, where
$\delta_{NR} = (15p-21)/10$ in the case of constant density medium,
and $p$ is the slope of the injected relativistic electron
distribution.
In our case, for any reasonable value of $p$, the transition should 
give  a {\it flattening} of the light curve, if it occurs after the jet break.

The time $t_{NR}$ of such a transition is not
unambiguously predicted theoretically, but 
for GRB 970508 ($z$=0.835) it  was seen to occur
at $t_{NR}\sim 100$ days
(Frail, Waxman \& Kulkarni 2000).  Since in that case 
an isotropic equivalent total kinetic energy of $E=5\times10^{51}$ erg and 
$n=0.5$ cm$^{-3}$ were estimated, 
we can assume 
\begin{equation}
t_{NR}\sim 55 (1+z) \left ({ E_{52} \over n}\right)^{1/3}  {\rm days.}
\end{equation}
This value is also consistent with theoretical expectations
(see e.g. Dai \& Lu 1999; Livio \& Waxman 2000; Huang, Dai \& Lu 1998) 
and also with the interpretation
given by \cite{berger03} of the radio light curve of GRB 980703.
With the parameters derived by \cite{berger} for the wide jet
component of GRB 030329, we obtain 
\begin{equation}
t_{NR}\, \sim\, 44\, \left( {E\over 5.7\times 10^{51} {\rm erg} }
\right)^{1/3}\left({1.8 \over n}\right)^{1/3}\, \, \, {\rm days,}
\label{tnr}
\end{equation}
which is close to the time of the first \xmm observation.
The flat time decay ($\delta=1.45\pm 0.2$) 
can thus be explained
by the wide jet component which becomes non--relativistic.
This value of  $\delta_{NR}$ 
gives  a slope of the electron distribution $p=2.37\pm0.13$,
corresponding to an  X--ray spectrum with $\Gamma=(p/2+1)=2.18\pm 0.07$.
The \xmm spectra are consistent with this 
value\footnote{the coincidence  with the
spectral slope derived by the first {\it Rossi--XTE}
observation is, in this scenario, a pure coincidence,
since the early X--ray emission is due  to the narrow jet.}.

At these late times, the cooling frequency is already below
the optical band. Therefore, the optical and the X--ray
light curves should have the same decay slope.
After accounting for the 
contribution from the supernova,
\cite{berger} and  \cite{lipkin} derived  $\delta_{opt}\sim$ 2.35 
between day 10 and day 40 (i.e. after the jet break of the wide jet).
This is in agreement with the expected value  $\delta_{opt}=p$
(\cite{sari}).
Note that the optical light curve after 40 days is dominated by the
supernova light, which masks the predicted flattening.

Finally, we would like to caution the reader that there may exist
other possible interpretations of the observed flattening
of the X--ray light-curve.
In the cannonball scenario (e.g. Dado, Dar \& De Rujula 2002) 
a flattening of the light curve can be 
associated to the afterglow contribution of a new cannonball: indeed,
in Dado, Dar \& De Rujula (2004) the optical afterglow of GRB 030329 is
explained with two cannonballs (plus the SN contribution) up to 70
days after the trigger.
A flattening of the X--ray light-curve could then be possible if 
a third cannonball starts to contibute in the X--ray band
(but not in the optical, which is dominated by the SN light).
However, the number of cannonballs should be equal to
the number of main pulses in the prompt
$\gamma$--ray emission, which are only two in GRB 030329.

In the dyadosphere model (e.g. Ruffini et al. 2003),
a flattening of the light-curve is predicted, as in the
more conventional fireball scenario, when a transition to
the Newtonian expansion occurs, but other features specific
to GRB 030329 (such as the break at $\sim 0.5$ days) would
remain unexplained.

\begin{figure}
\psfig{figure=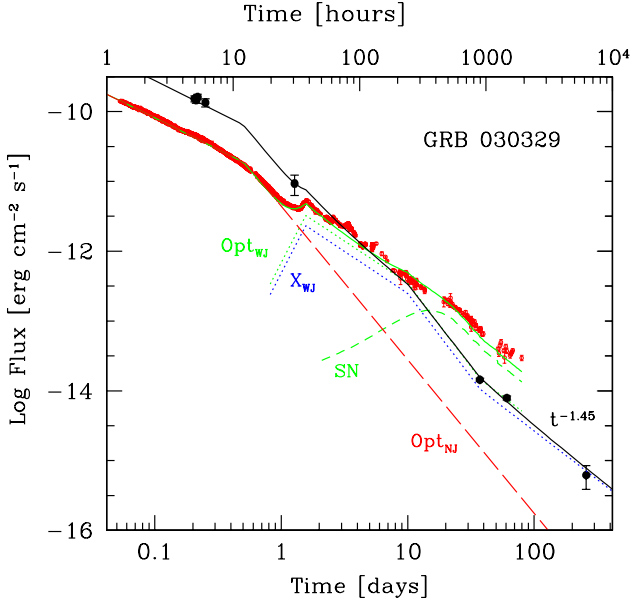,angle=0,width=9cm}
\vskip -0.5 true cm
\caption{Optical (Lipkin et al. 2004) and X--ray [0.5--2 keV] light curves of the 
GRB 030329 afterglow.
The two solid lines (one for the X--ray and one for the optical)
correspond to the prediction of 
the narrow+wide jet scenario,
modified at late times
to account for the non--relativistic phase of the wide jet.
Dotted lines indicate the contribution of the wide jet only,
the short--dashed line corresponds to the optical contribution of the 
supernova, and the long--dashed line corresponds to the optical 
contribution of the narrow jet, with a break at 12 hours.
The jet--break time for the wide component is assumed to occur
10 days after the burst.
}
\label{xmm_new}
\end{figure}

\section{Conclusions}

The combination of the exceptional brightness of \gr and the high
sensitivity of \xmm has allowed us to study a GRB X--ray afterglow up
to very late time, leading to a $\lsim$3$\sigma$
detection of the afterglow 258 days after the burst.

Although other interpretations are possible,
the \xmm data at $t=$37, 61 and 258 days are consistent with the X--ray 
light curve being dominated by the wide jet component envisaged by Berger et
al. (2003), Lipkin et al. (2004) and Sheth et al. (2003).  
In particular, and more importantly, we have evidence that this component
became non--relativistic at $\sim$40 days after the GRB explosion, in
remarkable agreement with what predicted assuming the parameters
(kinetic energy and density of the external medium) derived by the
above authors by fitting the radio and optical light curves.  The
immediate prediction of this interpretation is that also the radio light
curve should show a flattening from $\delta=2$ to
$\delta_{NR}\sim 1.5$ after $t\sim$40 days.
The publicly available radio data at these late times have large 
uncertainties and cannot exclude such a behavior. Only further analysis
of more radio data could settle this issue.
The lack of a  flattening in the radio light curve 
would imply that the fireball has not yet reached the non--relativistic
phase, due to 
a larger total kinetic energy of the wide jet and/or a smaller density
for the interstellar medium.
This would then require a  different reason for the
observed flattening in the X--ray light curve,
e.g. a larger supernova contribution in the X--ray band.
This would imply an X--ray luminosity
at least one order of magnitude higher than that  
observed in other type  Ic and  II supernovae 
(see Fig. 3 of \cite{kouveliotou}).

\begin{acknowledgements}
Based on observations obtained with \xmm, an ESA science
mission  with  instruments and contributions directly funded by
ESA Member States and NASA.
We thank the  \xmm Project Scientist Fred Jansen for granting 
time to observe this source and  Davide Lazzati for useful discussions.
This work has been supported by the Italian Space Agency and by  MIUR
(COFIN  2003020775).
\end{acknowledgements}

\end{document}